\documentclass[epj,final]{svjour}
\usepackage{graphicx}
\usepackage{psfrag}

\newcommand{\bq}{\begin{equation}}
\newcommand{\eq}{\end{equation}}
\newcommand{\ba}{\begin{eqnarray}}
\newcommand{\ea}{\end{eqnarray}}
\newcommand{\dd}{{\rm d}}

\begin{document}

\title{Power laws in zero-range processes on random networks}
\author{B. Waclaw\inst{1}\and Z. Burda\inst{2,3}\and W. Janke\inst{1,4}}
\institute{
$^1$Institut f\"ur Theoretische Physik, Universit\"at Leipzig,
Postfach 100\,920, 04009 Leipzig, Germany \\
$^2$Marian Smoluchowski Institute of Physics,
Jagellonian University, Reymonta 4, 30059 Krak\'ow, Poland \\
$^3$Mark Kac Complex Systems Research Centre, Jagellonian University, Krak\'ow,
Poland \\
$^4$Centre for Theoretical Sciences (NTZ), Universit\"at Leipzig, Emil-Fuchs-Stra{\ss}e 1, 04105 Leipzig, Germany}

\abstract{
We study statistical properties of a zero-range process (ZRP) on random networks.
We derive an analytic expression for the distribution of particles (also called node occupation distribution)
in the steady state of the ZRP in the ensemble of uncorrelated random graphs. 
We analyze the dependence of this distribution on the node-degree distribution.
In particular, we show that when the degree distribution is tuned properly, one can obtain
scale-free fluctuations in the distribution of particles.
Such fluctuations lead to a power law in the distribution of particles, 
just like in the ZRP with the hopping rate $u(m)=1+b/m$
on homogeneous graphs.
}
\date{19.02.2008}

\maketitle

\section{Introduction}
Many statistical systems are defined on random networks or random
lattices. Usually, this means that one considers a system on a lattice with some irregularities, or
even on a purely random network, but instead of looking at what happens on a single network, one performs an
annealed average over a statistical ensemble of networks that is over a collection of random graphs with some statistical weights.

The question of how averaging over the disorder influences the statistical properties of the system has been previously addressed in the context of two-dimensional statistical models. 
It was shown that additional degrees of freedom related to fluctuations of the geometry can
lead to quite distinct behavior, in comparison to analogous systems defined on a fixed lattice.
For example, critical 
properties of the two-dimensional Ising model on a fixed lattice 
are described by Onsager exponents \cite{onsager} while on a random lattice,
represented as a sum over planar networks, by ``dressed'' KPZ-DDK ones \cite{kpz,grav}.
A similar change of critical exponents is observed for other models.

In this paper we address the same problem but for a statistical
system on complex random networks. As a particularly simple but interesting example we consider
a zero-range process (ZRP), which has been thoroughly studied on fixed networks \cite{cc1,cc2,jdn,jdn2,jdn3,florence,zrp-long}.
The particular feature of this out-of-equilibrium model is that under certain conditions particles tend to condense on a single node.
Here we analyze the influence of the averaging over random networks on
the distribution of particles in the steady state.
We shall discuss an ensemble of networks 
with given probability distribution of degrees. To make
things as simple as possible we shall restrict the discussion
to uncorrelated networks with independent node degrees. 

The paper is organized as follows. In the next section 
we shall recall the definition of the ZRP on a network
and introduce quantities describing the steady state 
of this model. Then we shall consider the free ZRP process
being a particular class of ZRP, in which the hopping
rates do not depend on the distribution of particles but
only on the connectivity of the network. For this case
we will present in Secs.~3 and 4 an exact solution to the problem of how
the distribution of particles averaged over nodes depends on the node-degree distribution
and how to choose the latter one to obtain a scale-free
distribution of particles. In the following Sec.~5 we 
will discuss finite-size
effects observed in the distribution of particles
for some models of random networks like Erd\"os-R\'enyi
graphs. We will finish in Sec.~6 with some concluding remarks.

\section{Model description}
Consider the ZRP on a connected simple graph with $N$ nodes and 
a sequence $\vec{k} = \{k_1,\ldots,k_N\}$ of node degrees.
The state of the ZRP is given by the distribution of particles 
$\vec{m}=\{m_1,\ldots,m_N\}$ on nodes of the network, where $m_i\geq 0$ is the number 
of particles at node $i$. The total number of particles 
$M=m_1+\ldots+m_N$ is conserved during the process. 
A particle can hop from node $i$ to one of its neighbors with rate $u(m_i)/k_i$. 
The function $u(m)$, called hopping rate, depends only on the 
number of particles at the departure node $i$. 
The factor $1/k_i$ takes care of distributing the outflow of particles 
equally between the neighbors. The hopping rates $u(m)$ are non-negative
and identical for all nodes.

The ZRP is known to have a unique steady state \cite{evans}.
Static properties of this state are described by a partition 
function $Z(N,M,\vec{k})$ depending only on the degree sequence $\vec{k}$:
\bq
Z(N,M,\vec{k}) = 
\sum_{m_1,\dots,m_N=0}^M \delta_{m_1+\dots+m_N,M}
\prod_{i=1}^N p(m_i) k_i^{m_i},	\label{part}
\eq
with statistical weights $p(m)$ defined as
\bq
p(m)= \prod_{n=1}^m \frac{1}{u(n)}, \;\;\; p(0)=1.	\label{pbyu}
\eq
We will refer to $Z(N,M,\vec{k})$ as to a microcanonical 
partition function. The main quantity describing the system is 
the distribution of particles $\pi(m)$, averaged over all configurations $\vec{m}$
with the weight given by the partition function (\ref{part}) and over all nodes:
\bq
\pi(m,\vec{k}) = \frac{1}{N} \sum_i \langle \delta_{m,m_i}\rangle,
\eq
where the argument $\vec{k}$ means that it is calculated for a single network with a given sequence of degrees.
It is also called node occupation distribution.
It can be calculated \cite{zrp-long} as follows:
\bq
\pi(m,\vec{k}) = 
\frac{1}{N} \sum_{i=1}^N 
\frac{Z_i(N-1,M-m,\vec{k}')}{Z(N,M,\vec{k})} k_i^{m} p(m), \label{pigeneral}
\eq
where $Z_i$ is the partition function for a graph with $N-1$ 
nodes and degrees $\vec{k}'=\{k_1,\dots,k_{i-1},k_{i+1},\dots,k_N\}$. Equation (\ref{pigeneral}) holds for
any connected graph with a given degree sequence. 

Suppose now that we are interested in the behavior of the ZRP on 
a random network. In this case we have to take the average over networks. Denote by
$P(\vec{k})\equiv P(k_1,\ldots,k_N)$ the probability
of choosing a network with the degree sequence $k_1,\ldots,k_N$.
We can now define a canonical partition function
as the average over all degree sequences:
\bq
Z(N,M) =\sum_{k_1,\dots, k_N} P(\vec{k}) Z(N,M,\vec{k}) .
\label{canon}
\eq
In general, $P(\vec{k})$ may have a complicated form.
We shall restrict our attention to uncorrelated networks 
\cite{bk,homnasz} for which $P(\vec{k})$ is a product measure:
$P(\vec{k}) = \Pi(k_1)\cdots \Pi(k_N)$.
This means that node degrees are independent of each other
and that the observed degree distribution is $\Pi(k)$.\footnote{We neglect the fact that the total number of links is often fixed, which leads to an additional Kroenecker delta constraint $\delta_{2L,k_1+\dots+k_N}$. This constraint can be usually neglected in the thermodynamic limit.}
The canonical partition function
assumes then a simple, symmetric form:
\bq
Z(N,M) = \sum_{m_1,\ldots,m_N} \!\!
\delta_{m_1+\ldots+m_N, M} \, \prod_{i=1}^N \widehat{p}(m_i), \label{zzz}
\eq
where
\bq
\widehat{p}(m) = p(m) \sum_{k=1}^\infty \Pi(k) k^m
\eq
is an effective weight for a node occupied by $m$ particles.
As we see, the effective weight $\widehat{p}(m)$ is calculated
from the node degree distribution $\Pi(k)$ 
and the occupation weight $p(m)$. 
The effective partition function $Z(N,M)$ in Eq.~(\ref{zzz})
has the form of a partition function of the balls-in-boxes model 
with identical weights,
which has been thoroughly studied \cite{bbj,bb1,bbj2}. 
$Z(N,M)$ is invariant with respect to any permutation $\sigma$ of node 
occupation numbers $m_i \rightarrow m_{\sigma(i)}$. 

\section{Free ZRP}
In this section we shall consider a particular example of
a ZRP for which the  hopping rate $u(m)=1$ is independent of $m$. 
We shall call it free ZRP (FZRP). In this case, also
the occupation weight  (\ref{pbyu}) is constant, $p(m)=1$,
and the canonical partition function (\ref{zzz}) reduces to
\bq
Z(N,M) = \sum_{m_1,\dots,m_N} \!\!
\delta_{m_1+\ldots+m_N, M} \, \prod_{i=1}^N \mu(m_i), \label{zz}
\eq
where
\bq
\mu(m) = \sum_{k=1}^\infty \Pi(k) k^m
\label{mom}
\eq
is the $m$-th moment of the node-degree distribution.
The probability that a node is occupied by $m$ particles is now:
\bq
\pi(m) = \frac{Z(N-1,M-m)}{Z(N,M)} \mu(m),
\label{pisym}
\eq
just as it was in the balls-in-boxes model \cite{bbj}.

For further convenience, let us introduce a generating function for the moments $\mu(m)$:
\bq
M(z) = \sum_m \mu(m) \frac{z^m}{m!} ,
\label{mz}
\eq
which encodes the same information as $\Pi(k)$. Indeed, inserting
Eq.~(\ref{mom}) into the last equation we see that it can
be interpreted as a Fourier transform of the node-degree distribution,
\bq
M(-iz) = \sum_k \Pi(k) e^{-ikz} .
\eq
From the generating function 
one can formally reconstruct the moments,
\bq
\mu(m) = \frac{m!}{2\pi i} \oint M(z) z^{-m-1} dz,
\label{inv_trans}
\eq
as well as the degree distribution,
\bq
\Pi(k) = \frac{1}{2\pi} \int_{-\pi}^\pi dz\, e^{izk} M(-iz).	
\label{piviaMz}
\eq
The partition function (\ref{zz}) is well defined if all moments of the 
distribution $\Pi(k)$ are finite. 
Usually, we are interested in the behavior of the system
in the thermodynamic limit $N\rightarrow \infty$. 
We can distinguish two cases: (a) the limiting distribution
$\Pi(k)$ for $N\rightarrow \infty$ has all moments finite, as
for instance for Erd\"os-R\'enyi (ER) graphs, where it is Poissonian,
(b) higher moments of the limiting distribution $\Pi(k)$ diverge for
$N\rightarrow \infty$ 
as it happens for scale-free graphs \cite{dorog}. For (a), the large
$N$ limit presents no difficulty, while for (b) it has to be
taken very carefully since it depends on the details of how
the ensemble is defined. 
Moreover, in case (a) one can show that for random graphs\footnote{By a random graph we understand here a graph being maximally random among all graphs with a given sequence of degrees.}
the probability of any sequence of degrees $\vec{k}$ factorizes in the limit $N \rightarrow \infty$
\cite{dorog}. This factorization often breaks down for (b). One observes particularly strong deviations from 
the factorization for $\Pi(k) \sim k^{-\gamma}$ with $2<\gamma\le 3$ where 
finite-size effects are especially strong \cite{bk,extr2,myphd}.
Below we shall discuss only the case (a) which is free of these problems.

\section{Power-law distribution of particles}
We have shown in the previous section that averaging over
fluctuating geometries leads to an effective model 
with the partition function (\ref{zz})
and weights $\mu(m)$ being the moments of the node-degree distribution. This model has an interesting
critical behavior for weights which fall off like $\mu(m) \sim m^{-b}$. 
For example \cite{god}, for 
\bq
\mu(m) \propto \Gamma(m+1)/\Gamma(m+b+1) \sim m^{-b}, \label{pmb}
\eq
one observes a condensation of particles  
when the density of particles $\rho=M/N$
is larger than a critical density $\rho_c=1/(b-2)$. 
In the thermodynamic limit, at the critical point $\rho=\rho_c$, fluctuations of the number of particles
become scale-free and $\pi(m)=\mu(m)$ displays a power law.
Below $\rho_c$ it has an exponential cut-off:
\bq
\pi(m) = \mu(m) \exp(A-B m),
\eq
where the constants $A$ and $B$ are chosen so that \linebreak
the normalization $\sum_m \pi(m) =1$ and the density of particles
$\sum_m m \pi(m)=\rho$ are fixed.
Above $\rho_c$, the distribution $\pi(m)$ is approximately 
given by $\mu(m)$ but with an additional peak centered 
around $m_*\approx M-\rho_c N$.

One now can ask whether the weights (\ref{pmb}) can be obtained in our FZRP
by tuning the node-degree distribution of the underlying network. 
Before we proceed, it is important to notice that 
the model given by the partition function (\ref{zz}) 
is invariant with respect to the rescaling:
\bq 
\mu(m) \rightarrow {\cal N} \phi^m \mu(m).
\label{AB}
\eq 
Indeed, the partition function (\ref{zz}) changes only
by a factor: $Z(N,M) \rightarrow {\cal N}^N \phi^M Z(N,M)$,
which is constant for given $N$ and $M$, while physical
quantities stay intact because the normalization factor
cancels out. Thus, we expect that if the moments are given by
\bq
\mu(m) = 
{\cal N} \frac{\Gamma(m+1)}{\Gamma(m+1+b)} \phi^m,  \label{mugamma}
\eq
then the degree distribution of node occupation numbers
at the critical density should be given by 
\bq
\pi(m) \propto \frac{\Gamma(m+1)}{\Gamma(m+b+1)} \sim m^{-b}. \label{pmb1}
\eq
The question we face now is whether there is a node-degree distribution
$\Pi(k)$ which has moments given by Eq.~(\ref{mugamma}). First of all, we 
observe that the parameter $\phi$ in Eq.~(\ref{mugamma}) plays the role of a
scale parameter of the distribution $\Pi(k)$ as follows from 
the definition of the moments (\ref{mom}): under the 
rescaling $k\rightarrow k/\phi$ the moments change as 
$\mu(m) \rightarrow \mu(m) \phi^m$.
We will use the freedom of choosing the parameter $\phi$ to fix the average degree $\bar{k}$ and thus also the number of links $L=\bar{k} N/2$.
The parameter ${\cal N}$ has to be chosen in such a way
that $\Pi(k)$ takes the proper normalization of a probability.

\begin{figure}
\center
\psfrag{xx}{$k/\phi$} \psfrag{yy}{$\Pi(k/\phi)/\phi$}
\includegraphics*[width=8cm]{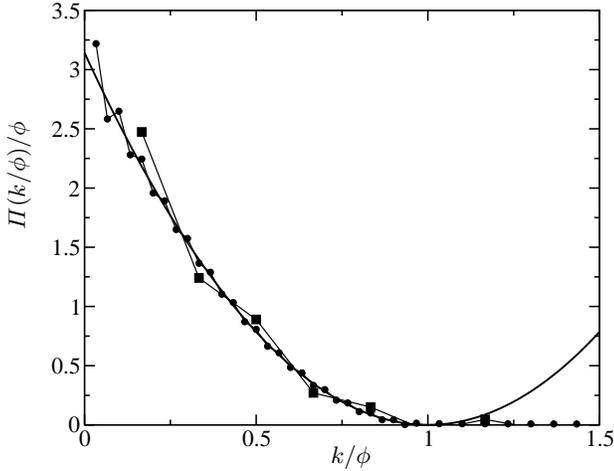}
\caption{$\Pi(k)$ calculated from the exact formula (\ref{piviaMz}) (points) and approximated one (Eq.~(\ref{piqgen}), thick line), for $\mathcal{N}=2\pi$ and $b=3$. Squares: $\phi=6$, circles: $\phi=30$. The approximate solution diverges for $x=k/\phi>1$ and has to be cut. For $0<x\leq 1$ the approximation is the better, the larger is $\phi$.}
\label{fig1}
\end{figure}

\begin{figure}
\center
\psfrag{xx}{$m$} \psfrag{yy}{$\mu(m)$}
\includegraphics*[width=8cm]{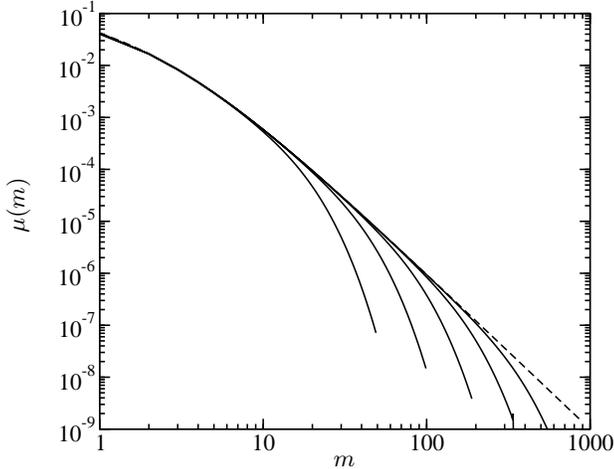}
\caption{Desired (dashed line) versus real distribution of particles $\mu(m)$ for networks with finite average degree $\bar{k}$ (solid lines), obtained from Eqs.~(\ref{piqgen}) and (\ref{mom}) for $b=3$.
Lines from left to right: $\phi=5,10,20,40,80$ which corresponds to $\bar{k}=1.67, 2.89, 5.4, 10.4, 20.4$ from Eq.~(\ref{phifor3}).
These plots approximate also $\pi(m)$ at the critical point. The parameter $\phi$ grows almost linearly with $\bar{k}$.}
\label{fig2}
\end{figure}

The integral in Eq.~(\ref{piviaMz}) is hard
to calculate and cannot be easily expressed 
in terms of elementary functions.
However, if we assume $\phi\gg 1$, then the function $M(-iz)$ goes 
to zero  sufficiently fast when $z\to\pm\infty$ and thus we can 
extend the limits of integration to $\pm\infty$. 
In this case the integral can be done analytically.
Equation (\ref{piviaMz}) becomes a Fourier transform of 
the function $M(-iz)$ which is a special case of the Mittag-Leffler 
functions having a known form of an infinite series expansion 
(see e.g. \cite{bw-is}). Changing variables $k\to x\phi$ we obtain
\bq
\Pi(x\phi) = \frac{\mathcal{N}}{2\pi\phi} 
\int_{-\infty}^\infty dz\, e^{izx} \sum_{m=0}^\infty 
\frac{(-iz)^m}{\Gamma(m+1+b)}.
\eq
According to Eq.~(B21) of Ref. \cite{bw-is}, the above integral yields
\bq
\frac{\mathcal{N}}{\phi} \sum_{m=0}^\infty \frac{(-x)^m}{m!\Gamma(b-m)},
\eq
and hence
\bq
\Pi(k) =  \frac{\mathcal{N}}{\phi} 
\sum_{m=0}^\infty \frac{(-k/\phi)^m}{m!\Gamma(b-m)} 
 = (\phi-k)^{b-1} \frac{\mathcal{N}}{\Gamma(b)\phi^b}.   \label{piqgen}
\eq
In Fig.~\ref{fig1} we compare $\Pi(k)$ computed numerically
using the integral (\ref{piviaMz}) with the original limit $\pm\pi$,
 and calculated by means of Eq.~(\ref{piqgen}).
Because the probability $\Pi(k)$ must be non-negative, the 
above solution is physical only for $k\leq \phi$ and we have to set $\Pi(k)=0$ for $k>\phi$. 
We see that the integer part of
$\phi$ can be interpreted  
as the maximal degree which can exist in the network. 
The existence of the upper cut-off in the node-degree distribution
is not only a property of the approximate solution. Also 
when one uses the exact relation~(\ref{piviaMz}) to calculate
the degree distribution $\Pi(k)$ for the moments of the form (\ref{pmb}),
one obtains negative values of $\Pi(k)$ for $k>\phi$, 
so again one has to cut off 
the solution and set $\Pi(k)$ to zero for $k>\phi$.
If one now calculates moments for the distribution (\ref{piqgen}) with 
the cut-off
directly from the definition (\ref{mom}), one will see that
they deviate slightly from those given by Eq.~(\ref{mugamma}).
However, the deviation decreases when $\phi$ increases and finally 
becomes negligible for sufficiently large $\phi$. 
In Fig.~\ref{fig2} we plot 
the moments $\mu(m)$ from Eq.~(\ref{mugamma}) and those
calculated from Eq.~(\ref{mom}), for various $\phi$. 
As $\phi$ increases, the curves tend asymptotically to a power law.

The parameter $\phi$ is related to the average degree as
$\bar{k}=\sum_{k=1}^\phi \Pi(k) k$.
For large $\phi$, the relation between $\phi$ and $\bar{k}$ is almost linear: 
\bq 
\bar{k} = \frac{\sum_{k=1}^\phi (k-\phi)^{b-1} k}{\sum_{k=1}^\phi (k-\phi)^{b-1}}
\approx \frac{\int_0^\phi (\phi-k) k^{b-1} \dd k}{\int_0^\phi k^{b-1} \dd k} 
 = \frac{\phi}{b+1}. \label{phi-to-k}
\eq  
This implies that in order to obtain the power-law distribution (\ref{pmb1}) for $N\to\infty$, the value of $\bar{k}$ should increase 
to infinity. For sparse networks $\bar{k}$ would be finite. We see thus
that the price to pay for having a scale-free distribution of particles 
is to make networks denser when their size increases. 

The normalization factor $\mathcal{N}$ must be chosen \linebreak 
so that the degree distribution is normalized to unity: 
$\sum_{k=1}^\phi \Pi(k)=1$. 
For example, for $b=2,3,4$ we obtain the following  
degree distributions $\Pi(k)$ for $0<k\leq\phi$:
\ba
  &b=2:& \;\;\; \frac{2(\phi-k)}{	\phi(\phi-1)}, \\
  &b=3:& \;\;\; \frac{(\phi-k)^2}{\phi(\phi-1)(2\phi-1)}, \\ 
  &b=4:& \;\;\; \frac{4(\phi-k)^3}{\phi^2(\phi-1)^2}, 
\ea
and zero for both $k=0$ and $k>\phi$, with $\phi$ given by the 
following formulas for $b=2,3$:
\ba
\phi &=&3\bar{k}-1, \\
\phi &=& \left(-1+4\bar{k}+\sqrt{1-16\bar{k}+16\bar{k}^2}\right)/2, 
\label{phifor3}
\ea
and by the solution of a cubic equation for $b=4$:
\bq
\frac{(1 + \phi)(3\phi^2-2)}{15\phi(\phi-1)} = \bar{k}.
\eq
How does it come about that the power laws are observed 
in the distribution of particles $\pi(m)$  
when one averages it over networks while they are not seen
in $\pi(m)$'s for individual nodes, for any single network in the ensemble? 
The answer is that the effective distribution 
$\pi(m)$ averaged over networks is a subtle result of a well-tuned
superposition of many exponential distributions: for a node with degree $k$, 
the distribution of particles is $\pi(m)\propto (k/k_{\rm max})^m$, where $k_{\rm max}$ is the maximal degree in the network \cite{florence}. On the node with maximal degree, however, there is a condensation just as for scale-free networks \cite{jdn3}, but it disappears in the thermodynamic limit. This happens because the critical density for the condensation becomes larger than $\rho=1/(b-2)$ which we assumed to hold in our system, and the system is always in the fluid phase.

\section{Other random graphs}
In the previous section we found a node degree distribution
for the ensemble of random, uncorrelated networks for which the corresponding
FZRP has a power-law particle distribution $\pi(m)$. What happens with FZRP on
generic random networks? Can the particle
distribution be scale free? Let us begin with what happens in
the limit $N\rightarrow \infty$. Consider some typical
examples of graphs like
random trees \cite{bbjk} or ER graphs for which the limiting
shapes of the degree distribution are known:
$\Pi(k) = e^{-1}/(k-1)!$ and $\Pi(k) =  e^{-\bar{k}}\bar{k}^k /k!$,
respectively. So we can calculate 
the corresponding critical
distribution $\pi(m)$.
In the first case, the generating function has a closed form
$M(z)=\exp(z+e^z)$, as follows from Eq.~(\ref{mz}), and
we can deduce the coefficients $\mu(m)$ from the inverse 
Laplace transform (\ref{inv_trans}). Using the saddle-point 
method and integrating around $z_0\approx \log(m/\log m)$ 
we can find the asymptotic behavior for large $m$:
\bq
	\log \mu(m) = m(\log m - \log\log m) + O(m).
\eq
We see that $\mu(m)$ grows over-exponentially for large $m$. 
This means that in the thermodynamic limit the condensation 
always happens, regardless of the density of particles. 
The distribution $\pi(m)$ in the bulk falls faster than any power law.
Similarly, one can estimate that for random ER graphs 
$M(z) \propto \exp(\bar{k} e^z)-1$ and the leading term 
in $\log \mu(m)$ is also $m\log m$. So again it is clear that in 
the limit $M\to\infty$ one cannot obtain a power-law distribution 
of particles. 

\begin{figure}
\center
\psfrag{xx}{$m$} \psfrag{yy}{$\pi(m)$}
\includegraphics*[width=8cm]{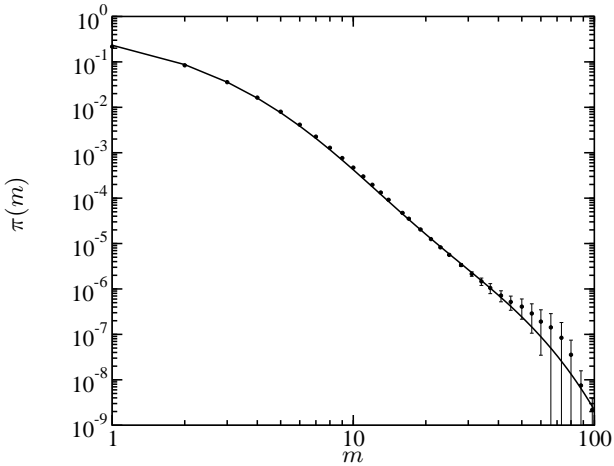}
\caption{The particle distribution $\pi(m)$ for ER graphs, $N=400$,
obtained from Eq.~(\ref{mom}) by using a Poissonian degree distribution
with $k_{\rm max} \approx 22$, calculated for $n=200$ samples
as described in the main text. The almost straight line on the 
log-log plot explains partially the quasi-power-law observed 
in numerical experiments (see the data points).
}
\label{fig4}
\end{figure}

Surprisingly, in Monte Carlo (MC) simulations 
of random trees and ER graphs one observes distributions $\pi(m)$ which
very much resemble those with power-law tails. Why does it happen?
The answer is that this is caused by finite-size effects.
When one repeats the calculations for finite systems, 
including sub-leading terms and a finite-size cut-off, one obtains a line like that in Fig.~\ref{fig4}, which very much resembles
a scale-free plot. The line is compared to MC simulations. 
In our analytic calculations aimed at mimicking the result of the simulation, finite-size effects were taken into account
as follows. In order to compute the moments we introduced a
finite-size cut-off to $\Pi(k)$.
The position of the cut-off $k_{\rm max}$ was estimated from 
the condition: $\Pi(k_{\rm max}) \approx 1/(nN)$, where $n$ is the number of samples in MC simulations. 
Next, for the distribution with the cut-off $k_{\rm max}$ we calculated the moments (\ref{mom}) and 
we got rid of the leading exponential behavior of $\mu(m)$,
using the freedom (\ref{AB}), by multiplying them by a factor 
$\exp(-m B)$ with $B$ appropriately chosen.
As we can see in Fig.~\ref{fig4} we obtained almost a 
straight line in the log-log plot. 
The discussion from the beginning of this section tells us, however, that this quasi-power-law behavior is only a finite-size effect 
which will disappear in the limit $N\rightarrow \infty$.

\section{Conclusion}
We analytically investigated the influence of annealed averaging over random networks
on the statistical properties of ZRP. In particular, we calculated how the particle distribution
depends on the node-degree distribution. We showed that by
tuning the node-degree distribution we can make the corresponding zero-range process
critical and having a power-law distribution of particles. 

We believe it is the first step towards the analysis of
more complex systems, where the topology and the dynamics 
of the system are coupled to each other 
and influence mutually. 
An example of this type of interactions was discussed in the context of
2d statistical systems on random lattices (2d gravity), where a back-reaction of the system on the lattice was observed \cite{kpz,grav,mwwj}, which manifested
as a change of fractal properties of the underlying geometry
when the system became critical.
It would be very interesting to see such an adaptation mechanism 
also for random complex networks.

\section*{Acknowledgments}
We thank the EC-RTN Network ``ENRAGE'' under grant No.~MRTN-CT-2004-005616
and the Alexander von Humboldt Foundation for support.
Z.~B. acknowledges support from a Marie Curie Actions Transfer of 
Knowledge project ``COCOS'',
Grant No. MTKD-CT-2004-517186 and a Polish Ministry of Science
and Information Society Technologies Grant 1P03B-04029 (2005-2008).
B.~W. thanks the DAAD for support.

\end{document}